\documentstyle[psfig]{caosp}

\begin{document}

%\pubyear{1993}
%\volume{23}
%\firstpage{7}

\htitle{The effects of convection on the colours of A and F stars}
\hauthor{B. Smalley and F. Kupka}

\title{The effects of convection on the colours\\ of A and F stars}

\author{B. Smalley \inst{1} \and F. Kupka \inst{2}}
\institute{Department of Physics, Keele University, Staffordshire ST5 5BG, UK \and
Institut f\"{u}r Astronomie, Universit\"{a}t Wien, Austria }

\date{December 18, 1997}
\maketitle

\begin{abstract}
We present a discussion on the effects of convection on the $uvby$
colours of A and F stars. The mixing-length theory used in {\sc ATLAS9}
is compared to the turbulent convection theory of Canuto \& Mazzitelli
(1991, 1992). Comparison with fundamental stars reveals that colours
calculated using the Canuto \& Mazzitelli convection theory are generally
in better agreement than those obtained using mixing-length theory.
\end{abstract}

\section{Introduction}

The colours calculated for stars later than mid A-type are affected by
treatment of convection. Small systematic errors were found in the
colours calculated using ATLAS6 (Kurucz, 1979), which could be due to
convection or missing opacity in the models (Relyea \& Kurucz, 1978).
Recent improvements in opacity in ATLAS9 models (Kurucz, 1991) ought to
ensure that opacity is now a less dominant source of the discrepancies. This
leaves convection as a possible source of the discrepancies.

We have compared mixing length theory (see Castelli et al., 1997) with
the turbulent convection model of Canuto \& Mazzitelli (1991, 1992). The
ATLAS9 code was used to calculate $uvby$ colours. The computations were
identical, except for treatment of convection. We considered three
cases:
\begin{enumerate}
\item mixing-length theory with approximate overshooting (MLT\_OV),
\item mixing-length theory without approximate overshooting (MLT\_noOV),
\item the Kupka (1996) implementation of the Canuto \& Mazzitelli theory (CM).
\end{enumerate}

Here we present a summary of the main findings from the comparisons
between the three treatments of convection outlined above. Full details
can be found in Smalley \& Kupka (1997).

\section{Comparison with Fundamental Stars}

The ultimate test of any model colours is to compare them to the colours
of stars whose atmospheric parameters have been obtained from direct,
model-independent methods. The fundamental stars presented by Smalley \&
Dworetsky (1995) were used to test the colours obtained from the three
grids.

We adopted the following methodology in the testing procedure: The
observed $uvby$ photometry of the fundamental star was de-reddened using
Moon's (1985) {\sc uvbybeta} programme. For each grid in turn, values of
$T_{\rm eff}$ and $\log g$ were obtained by interpolation within the
($b-y$, $c_0$) colour grid. We then compared either $T_{\rm eff}$ or
$\log g$ with the corresponding fundamental value. This way we could
compare $T_{\rm eff}$ and $\log g$ independently, since very few stars
have fundamental values for both parameters.

\begin{figure}
\psfig{figure=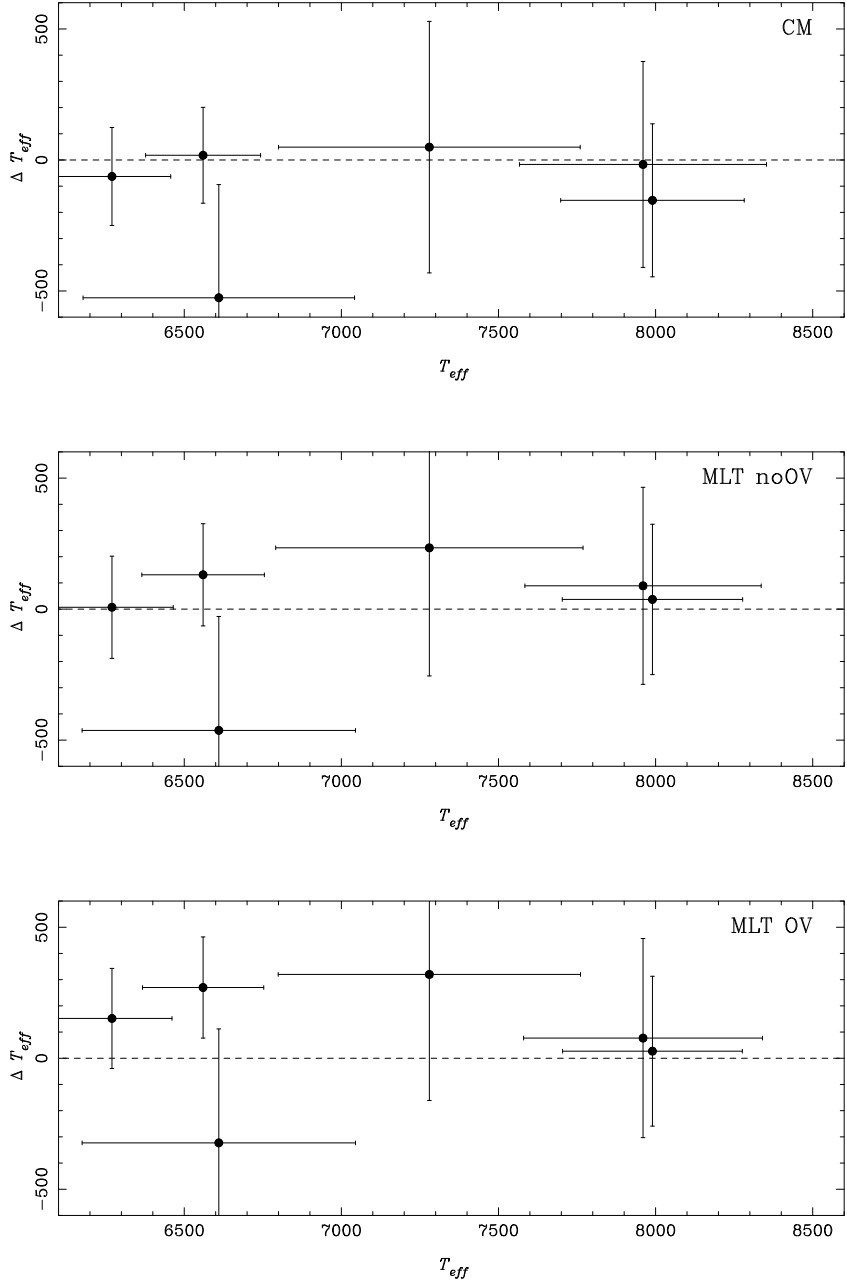,height=17cm}
\caption{Comparison of difference between grid and fundamental $T_{\rm eff}$ for
the 3 grids. $\Delta T_{\rm eff} = T_{\rm eff}({\rm grid}) - T_{\rm eff}({\rm fund})$.}
\end{figure}

Figure~1 shows the comparison between the fundamental $T_{\rm eff}$
values and those obtained from the 3 grids.
The CM model gives the best agreement. The MLT\_noOV grid is
somewhat discrepant, but still within the error bars.
The MLT\_OV grid is widely discrepant.

\begin{figure}
\psfig{figure=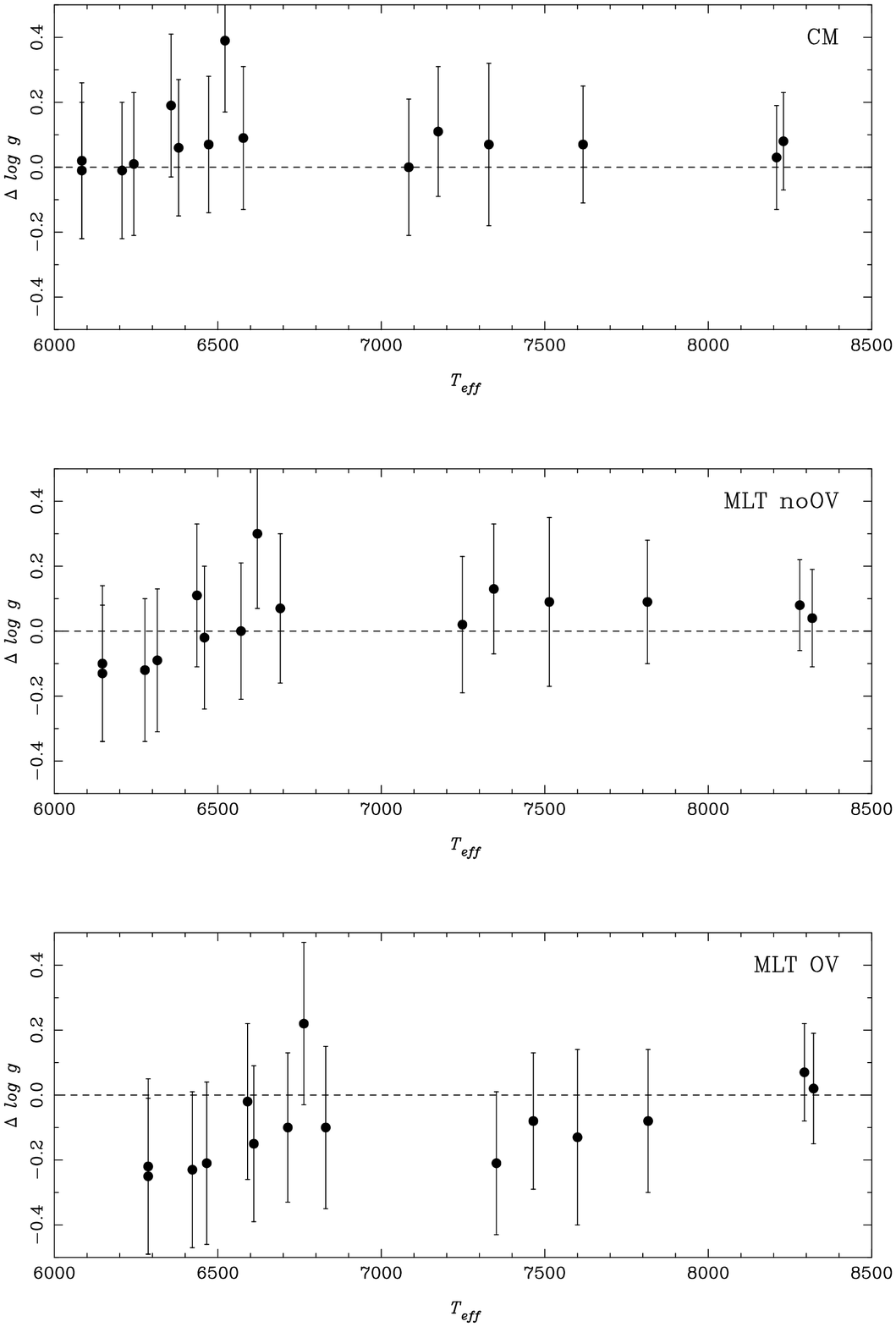,height=17cm}
\caption{Comparison of difference between grid and fundamental $\log g$ for
the 3 grids. $\Delta\log g = \log g({\rm grid}) - \log g({\rm fund})$.}
\end{figure}

Figure~2 shows the comparison between the fundamental $\log g$
values and those obtained from the 3 grids. There is little difference
between CM and MLT\_noOV grids, but MLT\_OV grid is clearly discrepant.
However, the CM models give less scatter and no hint of a possible
systematic trend at low $T_{\rm eff}$ observed in the MLT\_noOV models.

From the comparisons with fundamental stars we find that the CM grids
give the best agreement. The MLT\_noOV grids are only marginally less
successful in recovering the fundamental values. The MLT\_OV grids are
however significantly discrepant.

Since there are very few stars with fundamental values of $T_{\rm eff}$ or
$\log g$, the comparisons were also performed using non-fundamental
stars. Whilst this allows for significantly more stars in the
comparisons, there is a very real danger that model-dependent systematic
errors could bias any result. However, similar results were found using
stars which have values of $T_{\rm eff}$ obtained using the Infrared Flux
Method (Blackwell \& Lynas-Gray 1994), once known spectroscopic binaries
had been excluded. Open cluster stars can be used as surface gravity
standards by fitting evolutionary models to the cluster photometry. In
this case the stellar interior calculations can be influenced by the
treatment of convection (e.g. Stothers \& Chin 1995, Canuto et al. 1996).
Nevertheless, similar results were found for Hyades stars with $\log g$
obtained from evolutionary models.

Overall, the CM model has the greatest success in recovering the $T_{\rm eff}$
and $\log g$ obtained from both fundamental and non-fundamental methods.
However, the MLT\_noOV grids are only marginally less successful than the
CM grids. The MLT\_OV grids are clearly discrepant.

\section{Metallicity Effects}

The colours of late-A and F stars can be significantly affected by
metallicity, due to the vast amount of metal lines. The $m_0$ metallicity
index can be used to estimate [M/H] in late-A and F stars, usually via
empirical relations (Smalley 1993).

In all the three cases considered here, the theoretical $m_0$ values are
found to be discrepant. None of the model results based on the different
convection treatments agrees completely with the standard main-sequence
reference line, and all are discrepant for late-type stars. The problem
may not be due to convection alone, but could be related to
microturbulence (Smalley \& Kupka 1997), because microturbulence is
probably closely related to the small-scale part of the photospheric
convective flow (Holweger \& St\"{u}renburg 1993). Investigations into
the causes of this discrepancy are being undertaken.

\section{Conclusion}

From a comparison with the observed $uvby$ colours, we have found that
the CM grid gives results that are generally superior to those with MLT
theory without overshooting (MLT\_noOV). Models with overshooting
(MLT\_OV) are found to be clearly discrepant.

The metallicity index $m_0$ is not in agreement. The reason for this is
unclear, but could be linked to microturbulence.

\end{document}